\documentclass[11pt,english]{article}
\usepackage{mathptmx}

\usepackage[]{fontenc}
\usepackage[latin9]{inputenc}
\usepackage[letterpaper]{geometry}
\usepackage{amstext}
\usepackage{graphicx}
\usepackage{setspace}
\usepackage{subscript}
\makeatletter

\DeclareRobustCommand{\greektext}{%
  \fontencoding{LGR}\selectfont\def\encodingdefault{LGR}}
\DeclareRobustCommand{\textgreek}[1]{\leavevmode{%
  \IfFileExists{grtm10.tfm}{}{\fontfamily{cmr}}\greektext #1}}
\DeclareFontEncoding{LGR}{}{}
\DeclareTextSymbol{\~}{LGR}{126}

\newcommand{\lyxaddress}[1]{
\par {\raggedright #1
\vspace{1.4em}
\noindent\par}
}

\@ifundefined{date}{}{\date{}}
\makeatother

\usepackage{babel}
\begin{document}

\title{\textbf{Travelling Wave Magnetic Resonance Imaging at 3 Tesla }}

\author{\textbf{F. Vazquez, R. Martin, O. Marrufo, and A. O. Rodriguez}%
\thanks{Corresponding author: Alfredo O. Rodriguez, arog@xanum.uam.mx%
}}

\maketitle

\lyxaddress{\begin{center}
\textbf{Department of Electrical Engineering, Universidad Autonoma
Metropolitana Iztapalapa, Mexico, DF 09340, Mexico }
\par\end{center}}
\begin{abstract}
Waveguides have been successfully used to generate magnetic resonance
images at 7 T with whole-body systems. The bore limits the magnetic
resonance signal transmitted because its specific cut-off frequency
is greater than the majority of resonant frequencies. This restriction
can be overcome by using a parallel-plate waveguide whose cut-off
frequency is zero for the transversal electric modes and it can propagate
any frequency. To investigate the potential benefits of traveling-wave
excitation for whole-body imaging at 3 T, we compare numerical simulations
at 1.5 T, 3 T, 7 T, and 9 T via the propagation of the principal mode
of a parallel-plate waveguide filled with a cylindrical phantom and
two surface coils. B1 mapping was computed and used to investigate
the feasibility of the traveling-wave approach at 3T. The point spread
function method was used to measure the imager performance for the
traveling-wave magnetic resonance imaging experiment. Human leg images
were acquired to experimentally validate this approach. The principal
mode shows very little variations in the field magnitude along the
propagation direction at 3 T when compared to other higher resonant
frequencies. The B1 mapping showed that it is possible to conduct
experiments using the traveling-wave approach at 128 MHz. The point
spread function results showed a good performance of the scanner for
these type of experiments. Leg images were obtained with the whole-body
birdcage coil and the waveguide with two circular coils for comparison
purposes. The simulated and in vivo results correspond very well for
both magnetic field and specific absorption rate. A pretty similar
performance was observed for the traveling-wave approach and the conventional
one. We have demonstrated the feasibility of traveling-wave magnetic
for whole-body resonance imaging at 3T, using a parallel-plate waveguide
with standard pulse sequences and only one coil array. This extends
the use of the waveguide approach to a wider range of resonant frequencies.\pagebreak{}
\end{abstract}

\section{Introduction}

The use of travelling waves has been successfully implemented to generate
magnetic resonance images (MRI) at 7 T using whole-body systems {[}1{]}.
This approach allows one to cover samples with larger fields of view
using RF coils and a waveguide with different cross-sections. Much
work on traveling-wave MRI has been published in the Annual Meetings
of the International Society for Magnetic Resonance Imaging in Medicine
(ISMRM) and The ISMRM High Field Workshop in Rome since 2008 {[}2-5{]}.
From these results, it can be observed that the waveguide approach
has mainly been tested at 7 T with wide cylindrical bores measuring
approximately 60 cm in diameter for human applications. However, other
research groups have also investigate the feasibility of this new
approach at 9.4T {[}6{]}. The aim of this study was to extend this
approach to lower magnetic fields than 7 T. To the best of our knowledge,
this is the first attempt to use travelling wave MRI at 128 MHz with
a 60 cm magnet bore {[}7{]}. The cut-off frequency of cylindrical
waveguides imposes a limitation on the minimum diameter necessary
to transmit a magnetic resonance signal. A 7 T/65 cm MR system is
mandatory to transmit MR signals with a circular cross-section waveguide.
To overcome this limitation a parallel-plate waveguide (PPWG) was
employed because its cut-off frequency is zero for the lowest-order
transverse magnetic mode so all frequencies can propagate {[}8{]}.
Therefore, multinuclear experiments can be performed since a wide
spectrum of frequencies can be produced and detected. Simulations
of the B1 field were calculated at different resonant frequencies
using a simulated cylindrical phantom together with a PPWG with constant
cross-section and the results were compared. The Specific Absorption
Rate (SAR) was also computed to investigate RF safety. The simulation
configurations consisted of one RF coil located at one end of a waveguide
to excite an MR signal in the sample, and another one was placed at
the opposite end to receive the signal. The distance between these
two RF coils is much greater than the actual dimensions of the circular
coils. Circular coils were used for reception and transmission of
the MR signal in all simulations. There is still some controversy
to perform traveling-wave imaging experiments at 3T. A PPWG was built
and B1 mapping was carried out to show the feasibility of this approach
and, the optimal flip angle for transmission of the RF waves was done
too. In addition, the point spread function method was used to estimate
the performance of the imager for twMRI. To experimentally test the
viability of this approach, images of a healthy volunteer\textquoteright{}s
leg were acquired with the PPWG at 3 T in a clinical MR imager. This
imaging experiment was repeated using only a pair of RF surface coils
and a whole-body birdcage coil, which is usually embedded in the majority
of clinical MR imagers at 3 T and the results were compared.

\section{Theory}

To understand the propagation of modes in a PPWG, we studied their
mathematical description. Full derivations of the electromagnetic
equations of PPWG modes may be found in various textbooks {[}8-9{]}.
Waveguides may be thought of as hollow conductive metal pipes that
guide electromagnetic waves. They acts as directors of energy rather
than as signal conductors, in the normal sense of the word. The simple
waves that a waveguide can transmit are transverse electric (TE) waves
of various modes and transverse magnetic (TM) waves. In dealing with
transmission inside a hollow conductor, the waveguide consisting of
two perfect electrically conducting (PEC) plates is most likely the
simplest case. This waveguide offers a straightforward mathematical
solution and is commonly found in a number of physics and telecommunications
applications. A wave or transverse mode is a particular electromagnetic
field pattern of radiation propagated inside a waveguide. Two basic
types of transverse modes can be transmitted and occur because of
boundary conditions imposed on the wave by the waveguide. Transverse
modes are classified into different types: TE modes (Transverse Electric)
and TM (Transverse Magnetic) modes. Maxwell's equations must be solved
under particular boundary conditions to calculate the TE and TM modes
of a PPWG. 

From reference {[}8{]}, we then have$ $

\[
\frac{\partial^{2}}{\partial z^{2}}E_{y}-\frac{\partial^{2}}{\partial x^{2}}E_{y}=\omega^{2}\mu\varepsilon E_{y}\qquad(1)
\]

Eq. (1) describes the transversal electric (TE) modes. Similarly,
transversal magnetic TM modes can be derived, such that 

\[
\frac{\partial^{2}}{\partial z^{2}}E_{y}-\frac{\partial^{2}}{\partial x^{2}}E_{y}=-j\omega\mu\left(\frac{\partial}{\partial z}E_{x}-\frac{\partial}{\partial x}E_{z}\right)=\omega^{2}\mu\varepsilon H_{y}\qquad(2)
\]

To determine both the transversal magnetic and electric modes, it
is necessary to solve Eqs. (1) and (2).

\subsection{Parallel-plate waveguide TE modes}

The TE mode has its electric vector in a plane normal to the direction
of the propagation, but has a component magnetic field parallel to
the direction of propagation. The following boundary conditions were
applied: $E_{y}=0$ when $x=0$ and $x=a$, then 

\[
E_{y}=j\frac{E_{0}}{2}\left(\exp\left(-j\beta_{x}x\right)-\exp\left(j\beta_{x}x\right)\right)\exp\left(-j\beta_{z}z\right)=E_{0}\sin\left(\beta_{x}x\right)\exp\left(-j\beta_{z}z\right)\qquad(3)
\]

where $E_{0}$ is an arbitrary constant. If we assume a perfectly
conducting conductor with the boundary conditions above, $\sin\left(\beta_{x}a\right)=0$,
so $\beta_{x}a=m\pi\text{\textgreek{b}}\text{}$, $m=1,2,3\ldots$
Because $m\neq0$, this implies that the electric field is uniform
and tangent to the waveguide plates. The wave constant of propagation
is

\[
\beta^{2}=\beta_{x}^{2}+\beta_{z}^{2}=\omega^{2}\mu\epsilon\qquad(4)
\]

From Eq. (4) the propagation constant along \emph{z} is 

\[
\beta_{z}=\omega^{2}-\left(\frac{m\pi}{a}\right)^{2}\qquad(5)
\]

Now, for $\beta\neq0$ waves can propagate longitudinally in a PPWG,
for $\beta>0$ waves will run parallel along the waveguide plates,
and for $\beta<0$ waves rapidly decay and will not propagate, and
the modes are called evanescent modes. Eq. (5) determines the \emph{critical}
or \emph{cut-off} frequency, which forms the limit between the transmission
and attenuation regions: 

\[
f_{cut-off}=\frac{m}{2a\sqrt{\mu\epsilon}}\qquad(6)
\]

for $m=0,12,3\ldots$

\subsection{Parallel-plate waveguide TM modes}

The transverse magnetic mode is an entirely longitudinal electric
field with an axial component of the magnetic field

\[
H_{y}=\frac{H_{0}}{2}\left(\exp\left(-j\beta_{x}x\right)-\exp\left(j\beta_{x}x\right)\right)\exp\left(-j\beta_{z}z\right)=H_{0}\cos\left(\beta_{x}x\right)\exp\left(-j\beta_{z}z\right)\qquad(7)
\]

where $H_{0}$ is an arbitrary constant. Assuming the boundary conditions:
$x=0$ and $H_{y}=H_{0}$ and, at $x=a$, $\cos\left(\beta_{x}a\right)=1$
then 

\[
\begin{array}{cc}
\beta_{x}a=m\pi, & m=0,1,2\ldots\end{array}\qquad(8)
\]
 where $m=0$ implies that a uniform magnetic field tangent to the
plates can be formed and is called the Magnetic Transverse (TM) mode.
Consequently, the $f_{cut-off}$ (Eq. (6)) becomes zero. The $\mathrm{TM_{00}}$
mode is the TEM mode and the dominant mode of the waveguide and it
can propagate at any frequency.

\section{Numerical simulations }

The finite-element method (FEM) is a useful and powerful computational
technique for simulating electromagnetic waves that propagate and
interact with their surroundings. This method was used due to its
ability to model complex geometries with acceptable accuracy. FEM
is able to determine propagation characteristics such as loss and
dispersion and electromagnetic field distributions as a function of
spatial location, frequency, or time. The three-dimensional simulation
domain is discretized into tetrahedral mesh elements. To properly
simulate wave propagation, the largest mesh element size must be smaller
than $\frac{{\displaystyle cf_{0}^{-1}}}{3}$ {[}10{]}, where \emph{c}
is the speed of light and $f_{0}$ is the resonant frequency. The
experiment was simulated using a time-harmonic solver, such that only
one input frequency was considered at a time. The model problem was
solved using an iterative generalized minimal residual (GMRES) iterative
solver with symmetric successive over relaxation (SSOR) matrix preconditioning
{[}11{]}. 

All numerical simulations were performed using COMSOL MULTIPHYSICS
(V. 3.2, Comsol, Burlington, MA, USA). The TEM was computed at the
following frequencies: 128 MHz, 300 MHz, and 400 MHz (proton resonant
frequencies at 3 T, 7 T, and 9 T), along the \emph{z}-direction using
the software above which uses the equations {[}12{]}: 

\[
\begin{array}{c}
\nabla\times\left(\mu_{r}^{-1}\nabla\times E\right)-k_{0}^{2}\left(\epsilon_{r}-j\frac{\sigma}{\omega\epsilon_{0}}\right)E=0\qquad(9)\\
\\
\nabla\times\left(n^{-2}\nabla\times H_{n}\right)-\left(\beta_{r}^{2}n^{-2}-k_{0}^{2}\right)H_{n}=0\qquad(10)
\end{array}
\]

where, $\mu=4\pi\mathrm{x10^{-7}\, H/m}$, $\mu_{r}$ is the relative
permeability, $\epsilon=\epsilon_{0}\epsilon_{r}$, $\epsilon_{r}=1\mathrm{x}10^{-9}/36\pi\,\mathrm{F/m}$,
$\epsilon_{r}$ is the relative permittivity and $\omega$ is the
frequency. \emph{n} represents the refraction index inside the waveguide,
the propagation constant is $k_{0}^{2}=k_{x}^{2}+k_{y}^{2}+k_{z}^{2}$
and for $k_{z}=\beta$ represents the propagation constant along the
waveguide. To run all simulations, PEC plates were assumed and $p\times E=0$
(vanishing tangential electric field components at the shield surface). 

Two circular surface coils were used for reception and transmission
and were located at either end of the waveguide as shown in Fig. 1.a).
The surface coils were linearly driven. The waveguide plates were
mounted on an cylinder (diameter = 30 cm, 60 cm long) and assumed
to be 25 cm in width.

{\footnotesize }%
\begin{minipage}[t]{1\columnwidth}%
\noindent \begin{center}
{\footnotesize \includegraphics[scale=0.3]{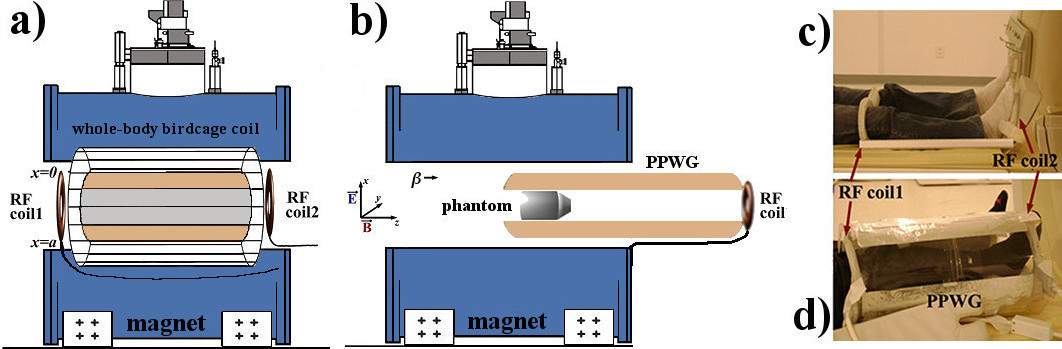}}
\par\end{center}{\footnotesize \par}

\begin{singlespace}
\begin{flushleft}
{\footnotesize Figure 1. Experimental setups using the PPWG with aluminum
plates for; a) simulation experiments using the PPWG inside the magnet
bore filled with a cylindrical phantom and, b) determination of the
}\emph{\footnotesize PSF}{\footnotesize{} showing boundary conditions
and the TEM mode propagation direction, $\beta$ (Eq. 5) using a pair
of RF surface coils: the distance between the phantom and the coil
was 150 cm, c) Photo showing leg and RF coils positions for image
acquisitions with pair of RF coils, and d) Photo of leg position and
location inside the PPWG. }
\par\end{flushleft}\end{singlespace}
\end{minipage}{\footnotesize \par}

\bigskip{}
The plates were made out of aluminum sheets and 1 V was applied in
all simulations. Both coils were tuned to 128 MHz and modeled using
impedance boundary conditions according to the software-provider company.
The excitation port was driven at the same resonance frequency. These
computations also included simulated phantoms: a) cylindrical phantom
($\sigma=1.5054\,\mathrm{S/m}$ and $\epsilon_{r}=69.062$), and b)
a stylized simulation phantom of a male was developed using simple
geometrical figures and based on the work by Fisher and Snyder {[}13{]}
and Kerr et. al. {[}14{]}. Our phantom was divided into 3 important
regions: 1) a sphere representing the head, 2) a box and a trapezoid
representing the chest and abdomen respectively, and 3) 8 ellipsoids
representing lower and upper extremities. These geometrical structures
were joined together to create the whole-body phantom. This phantom
was entirely designed with the COMSOL graphical tool. Fig. 4.a) shows
a schematic of the phantom. FEM simulations of the SAR were also calculated
for the same simulation setup.

\section{Waveguide prototype }

To test the viability of this traveling-wave approach at 3 T, two
PPWG prototypes were built: a) aluminum foil (6 \emph{\textgreek{m}}m
thickness) was laid on polycarbonate sheets separated by a 50 cm gap
and the waveguide plates were 45 cm wide and 160 cm long, and b) an
acrylic cylinder and two aluminum strips (25 cm wide and 60 cm long
and 6 \emph{\textgreek{m}}m thickness): strips were equally spaced
and mounted on the cylinder (30 cm diameter and 60 cm long). To avoid
interference with the field gradients, plates of both prototypes have
the same thickness. The the mechanical properties of the first prototype
allows us to have constant cross-section for this particular length
(Fig. 1.b)). It was used to test both the feasibility of travelling-wave
approach and performance of the MR scanner. The second prototype was
dedicated only to acquire images of a human leg. PPWG prototypes were
used together with a whole-body birdcage (68 cm long, 66 cm diameter
and 16 rungs and quadrature driven) for RF transmission, and reception
was performed with a vendor-provider coil array (Philips Medical Systems,
Best, NL) with two circular-shaped elements (11 cm diameter). Fig.
1.d) shows a photograph of the second waveguide prototype. The received-only
circular coils were positioned at both ends of the cylinder. Fig.
1.a) shows an illustration of the waveguide and the experimental setup.

\section{Methodology}

All imaging experiments were carried out on a 3 T Philips imager (Philips
Medical Systems, Best, NL) and operated under conventional mode for
clinical practice. Standard gradient echo pulse sequences were also
used in all cases. The first PPWG prototype as used for the optimal
flip angle, B1 mapping and PSF experiments and and according to the
experimental setup shown in Fig. 1.b). The second prototype was used
for the human leg imaging experiments only and the experimental setups
of Fig. 1.a).

\subsection{Optimum flip angle and B1 mapping }

Gradient echo sequences are normally used for twMR imaging but no
reference is given as to the optimal flip angle to be used {[}3-4{]}.
To experimentally determine the optimal flip angle for transmission
of the RF waves of the twMRI experiments with the PPWG, the image
SNR was computed using phantom images acquired at different flip angles.
T1-weighted images were acquired from 15\textsuperscript{0} to 90\textsuperscript{0}
with a 15\textsuperscript{0} step. Imaging experiments were conducted
with the following parameters: TR/TE = 1500/7.6 ms, FOV = 240x240
mm, matrix size=288x224, slice thickness = 5 mm, NEX = 1.

Accurate spatial information can be obtained via the computation of
B1 field mapping. As an attempt to solve the controversy of travelling-wave
MRI at 3T, we performed B1 mapping. Imaging experiments were conducted
using a PPWG, and 8-coil array for reception and a whole-body birdcage
coil for transmission at 128 MHz. A spherical phantom was located
at the magnet isocentre and the reception coil array was 100 cm away
as in the previous experiment. Images were acquired with the parameters
above. B1 maps were then calculated according to the method reported
in {[}15{]} and, a reference T1-weighted image was also acquired using
TR=350 ms, NEX=4 and the rest of the parameters above.

\subsection{Imager performance}

We used the point spread function (\emph{PSF}) to measure performance
of a 3T imager when used with a PPWG for travelling-wave MRI experiments.
The \emph{PSF} characterizes the response of an imaging system to
a point source. This parameter is able to predict of how the object
will be imaged, assuming a linear imaging process. The output image
may then be regarded as a two-dimensional convolution of the ideal
image with the \emph{PSF}: 

\[
Image_{output}=Image_{ideal}\star PSF\qquad(11)
\]

T1-weighted images were acquired for the \emph{PSF} experiments with
the following acquisition parameters: Flip angle = 20\textsuperscript{o},
TR/TE = 336.9/16.1 ms, FOV = 450 mm x 190 mm, matrix size = 500x169,
slice thickness = 10 mm, NEX = 5. Images were Fourier transformed
for the convolution in Eq. (11) and finally calculate the image \emph{PSFs}.
Two imaging experiments were performed to measure performance of a
PPWG in twMRI experiments and for comparison purposes. Images were
acquired with an 8-coil array operated in reception only, and transmission
was carried out with the whole-body birdcage coil for: a) standard
imaging experiments, and b) the PPWG and the surface coil. The first
experiment was intended for comparison purposes.

\subsection{Travelling-wave imaging experiments }

All imaging experiments were conducted according to the Bioethical
Committee of The Universidad Autonoma Metropolitana Iztapalapa. The
healthy volunteer gave his permission and signed an informed consent
form. The waveguide is long enough and its diameter wide enough to
accommodate an entire human leg. The volunteer's leg (height=1.54
m, weight=54 Kg) was 75 cm long and located in the waveguide as shown
in Fig. 1.d) to acquire images. T1-weighted images of an entire leg
were acquired for three cases: a) the whole-body birdcage coil for
transmission and reception, b) a pair of coil for reception together
with the whole-body birdcage coil for transmission, and c) the PPWG
and the pair of surface RF coils for reception and the whole-body
birdcage coil for transmission. The birdcage coil was quadrature-driven,
and the reception coils were linearly driven. The following acquisition
parameters for the three experiments were used: flip angle = 80\textsuperscript{0},
TR = 500 ms, TE = 4.6 ms, FOV = 484x171 mm, matrix size = 840x238,
slice thickness = 5 mm, NEX = 10.

\section{Results and Discussion }

From Eq. (6), the cut-off frequency $\left(f_{cut-off}=0\right)$
for the principal mode of the PPWG allows the transmission of all
the MR signals. This property of the PPWG offers the advantage to
be used in MRI applications with larger fields-of-view for different
magnet bores. The numerical computations of the magnetic fields of
a PPWG using the TEM along the \emph{z}-direction and different Larmor
frequencies were computed using FEM, as shown in Fig. 2. 

\begin{minipage}[t]{1\columnwidth}%
\noindent \begin{center}
\includegraphics[scale=0.4]{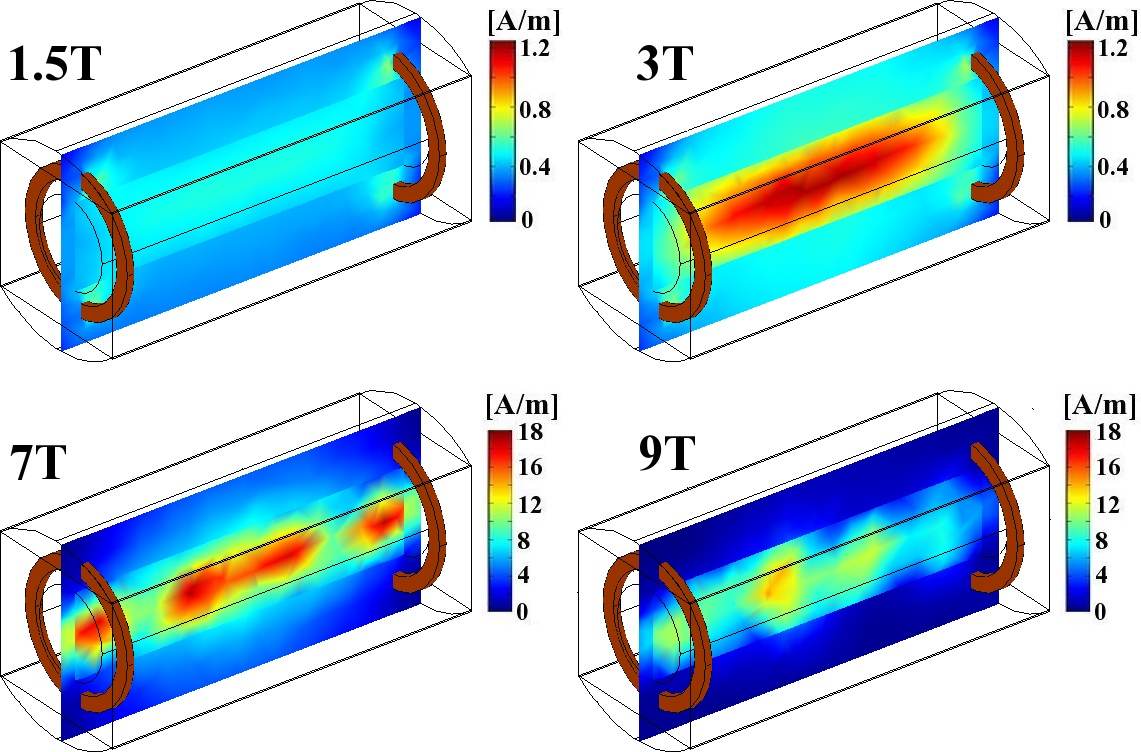}
\par\end{center}

\begin{singlespace}
\noindent \begin{center}
{\footnotesize Figure 2. FEM numerical simulations of B1+ distributions
for the setup in Fig. 1.a) at different magnetic field magnitudes.
A substantial increment in the B1+ magnitude can be appreciated as
the resonant frequency increases. However, the field pattern does
not seem to be the same: the 1.5T simulations shows a pretty uniform
behavior, contrary to the other simulations. }
\par\end{center}\end{singlespace}
\end{minipage}

\bigskip{}

A clear concordance can be seen with those simulations reported by
Brenner and co-workers {[}5{]}. An increment in the magnitude of B1
with the field strength can be observed. The numerically acquired
magnetic fields at 7 T and 9 T show magnitudes in the same range.
A very similar situation is obtained for the 1.5 T and 3 T cases.
A maximum theoretical 10-fold increase in the magnetic field strength
was numerically obtained for the 7 T and 9 T simulations compared
to those simulations at lower field intensities under the same conditions.
Field-uniformity plots were computed using the simulation data of
Fig. 2 and are shown in Fig. 3. All distribution magnetic field profiles
were calculated along the gray line as indicated in Fig. 3.b). The
B1 field distributions are reasonably smooth with a fairly analogous
pattern, although for the 3 T case, a better field uniformity can
be observed. The B1 field at 3 T shows a smoother uniformity and almost
no attenuation along the length of the waveguide. Meanwhile, for the
1.5 T TE mode simulation, B1 constant distribution with a significant
decrease can be observed. At high field, the 9 T simulation shows
a reasonable variation in the B1 magnitude and does not suffer much
attenuation along the propagation of the TE mode. These results are
very encouraging because the waveguide was dielectrically loaded and
represented by the cylindrical phantom and the resulting 3 T field
uniformity is very good. 

\bigskip{}

\begin{minipage}[t]{1\columnwidth}%
\noindent \begin{center}
\includegraphics[scale=0.3]{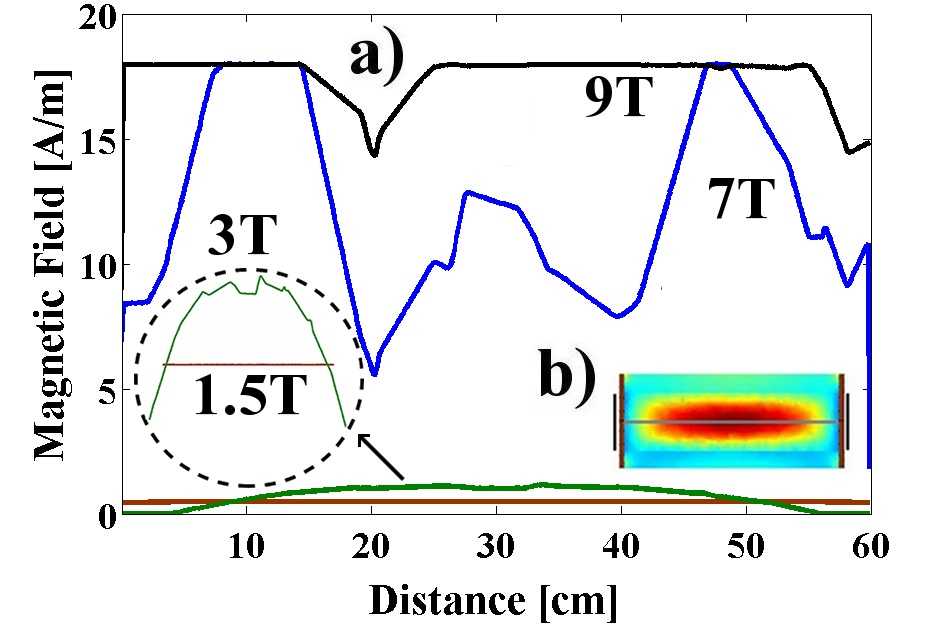}
\par\end{center}

\begin{singlespace}
\noindent \begin{center}
{\footnotesize Figure 3 a) Comparison plot of B1+ field distributions
along the waveguide for the field simulations of Fig. 2. The 1.5T
B1+ magnitude is constant a long the waveguide, while the other frequencies
show a clear increment towards the centre. The 7T and 9T profiles
tend to have a pretty similar magnetic field magnitude. b) All profiles
were taken along the grey line. }
\par\end{center}\end{singlespace}
\end{minipage}

\bigskip{}
The FEM-simulated B1 distribution of a simulated leg for the principal
mode was also acquired and is shown in Fig. 4.b). Despite the fact
that a simulated human leg representing a more complex condition was
introduced in the simulations of B1 fields (Fig. 4.a)), by simple
inspection the same pattern as that for the cylindrical phantom simulation
is obtained for the magnetic field. From Figs. 3.a) and 4.a), an increment
in the B1 field intensity at the very centre of the waveguide for
both 3 T simulations can be observed. This is related to the frequency
of the transmitted RF waves rather than the waveguide itself.

\bigskip{}

\begin{minipage}[t]{1\columnwidth}%
\noindent \begin{center}
\includegraphics[scale=0.6]{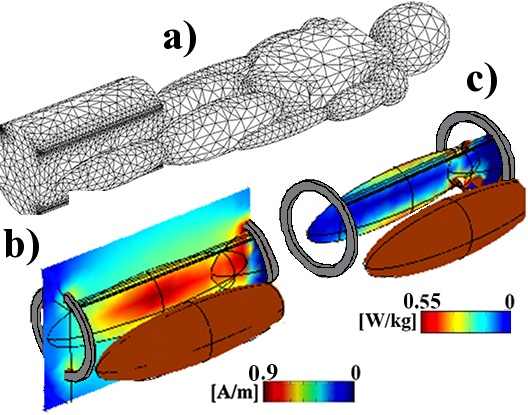}
\par\end{center}

\begin{singlespace}
\noindent \begin{center}
{\footnotesize Figure 4. Simulated magnetic field distribution obtained
at 3T for the TEM mode and SAR pattern of a phantom leg. The TEM simulation
shows a higher magnitude at the centre of the leg, and the magnitude
decreases towards the ends of the PPWG. A strong signal along the
the periphery and weak signal otherwise. A similar pattern to the
one in the phantom simulation (Fig. 2) was found for B1 simulations
at 3T. SAR values are within accepted limits for humans. }
\par\end{center}\end{singlespace}
\end{minipage}

\bigskip{}
Additionally, the distribution of the simulated B1 field for the simulated
human leg was also determined and is shown in Fig. 5.a). A similar
peak was produced by the waveguide and the leg at the waveguide centre,
as in the previous numerically acquired calculations for the cylinder
simulations. \bigskip{}

\begin{minipage}[t]{1\columnwidth}%
\noindent \begin{center}
\includegraphics[scale=0.2]{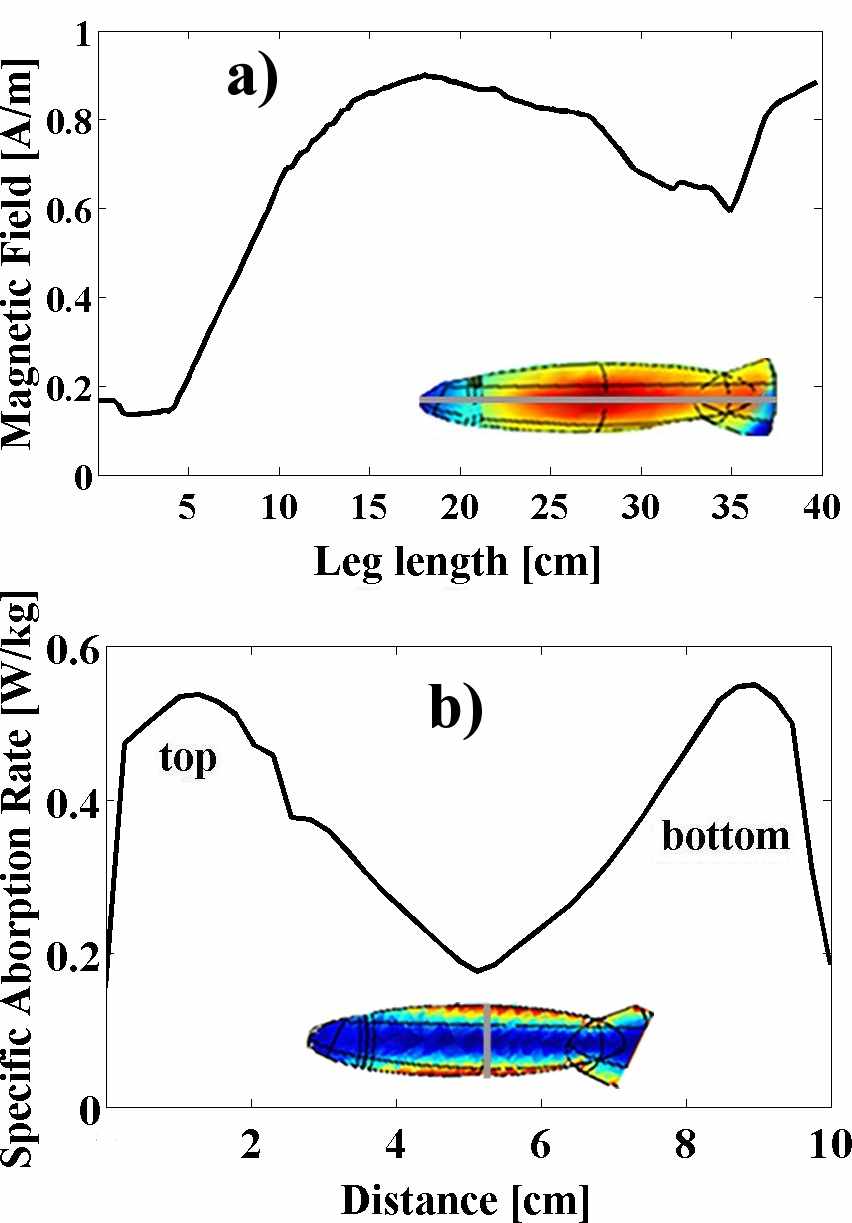}
\par\end{center}

\begin{singlespace}
\noindent \begin{center}
{\footnotesize Figure 5. a) Distribution of magnetic field along the
waveguide filled with the simulated leg of Fig. 4.b). b) SAR distribution
computed from simulation data of Fig.c): the two peaks represent the
hotspots in the simulated leg. Distributions were taken along the
grey lines in the simulated leg for each case. }
\par\end{center}\end{singlespace}
\end{minipage}

\bigskip{}
The FEM simulations were also used to verify the RF safety via the
SAR and the simulation setup of Fig. 1.a) with the simulated human
leg. Fig. 4.c) shows the SAR map of the simulated human leg. From
this, SAR hotspots are formed along the periphery and observed on
the anterior and posterior faces of the simulated leg, where SAR reaches
its maximum of 0.55 W/kg. An SAR distribution profile was also computed.
Fig. 5.b) shows the graph of the SAR distribution taken from the simulation
data of Fig. 4.c). However, although this numerical approach is limited,
because only simple geometrical figures were used to build the human
phantom, it still can be of some importance to understand the mechanisms
involved in the formation of the magnetic fields generating the MR
images with the PPWG for various Larmor frequencies. 

T1-weighted images of a spherical phantom using the experimental arrangement
of Fig. 1.b) were acquired for different flip angle values to determine
the optimal flip angle for RF wave transmission. Figs. 6.a)-f) show
the phantom images so from this image data, a plot of flip angle-vs.-SNR
was calculated and shown in Fig. 6.g). From this plot, it can be appreciated
that the maximum image SNR is reached at $\pi/2$. Therefore, a flip
angle around this value should be adequate for these type of travelling-wave
experiments.

\bigskip{}

\begin{minipage}[t]{1\columnwidth}%
\noindent \begin{center}
$ $\includegraphics[scale=0.3]{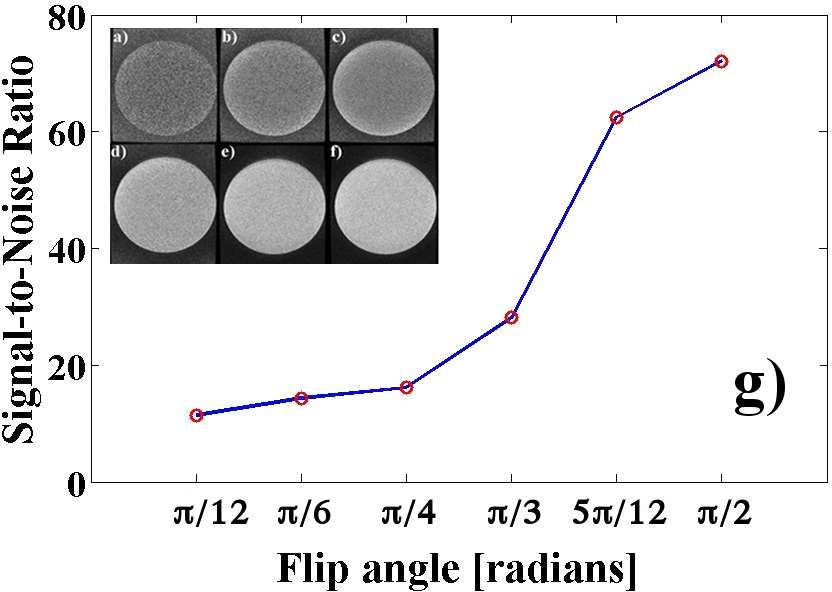}
\par\end{center}

\begin{singlespace}
\noindent \begin{center}
{\footnotesize Figure 6. T1-weighted phantom images were computed
for various flip angles (a)-f)) and a plot of flip angle-vs.-SNR (g)
was computed using the experimental configuration of Fig. 1.b). An
increment on the SNR can be observed as the flip angle increases until
founds its maximum at $\pi/2$. These results corresponds very well
with the standard gradient echo sequences.}
\par\end{center}\end{singlespace}
\end{minipage}

\bigskip{}

Previous to compute the B1 mapping, an attempt to acquire images without
the PPWG was performed and no signal was detected by the coil array.
We think the RF waves have no mechanical conducting instrument to
travel along to reach the distant object. A reference image was acquired
with the coil array under standard conditions giving a SNR of 20.65,
using an ROI of 70\% of the total volume. Standard deviation was 5\%
below the mean signal intensity. The normalized B1 map and a profile
of the field homogeneity were computed with the images acquired for
the flip angle experiments. The reference and B1 map images are shown
in Fig. 7.a) and b), respectively. the B1 map has a very good concordance
with those reported in {[}3{]}. A homogeneity profile was also calculated
with the Fig. 7.b) map and plotted in Fig. 7.c). The B1 map profile
shows an excellent uniformity despite the fact that the images were
acquired 1m away from the coil array and outside the imager. 

\begin{minipage}[t]{1\columnwidth}%
\noindent \begin{center}
\includegraphics[scale=0.3]{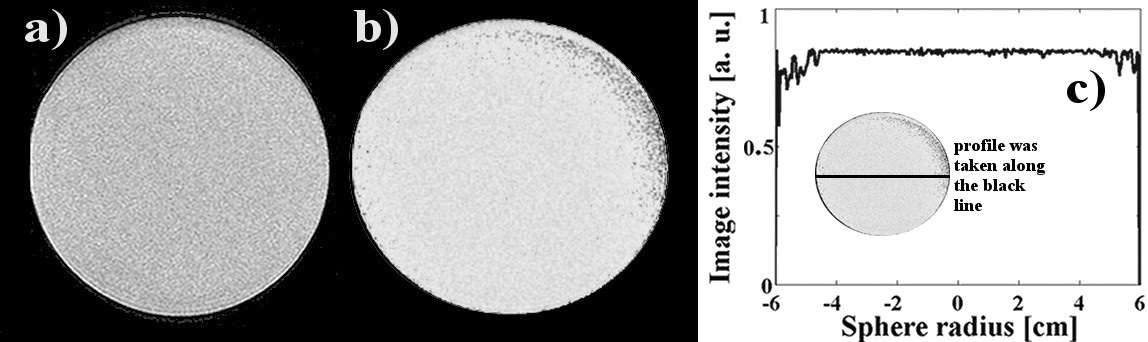}
\par\end{center}

\begin{singlespace}
\noindent \begin{center}
{\footnotesize Figure 7. Reference T1-weighted image of a spherical
phantom (a) acquired at 128 MHz and, the B1 map (b) and its corresponding
uniformity profile (c) were computed. According to the uniformity
profile in (c) the PPWG is able generate images with good quality.
The RF waves were transmitted from the negative z-direction according
to the experimental setup in Fig. 1.b). }
\par\end{center}\end{singlespace}
\end{minipage}\bigskip{}
This demonstrates that images generated with the PPWG have a good
image quality compared to the standard image acquisition methods.
It is important to highlight that the imager was operated using standard
clinical protocols to acquire images and no modifications was done
to the system. 

Also, T1-weighted images were acquired for the evaluation of the imager
performance using Fig. 1.b) configuration is shown Fig. 8.c). Fig.
8.b) shows an image obtained with the coil array and under standard
conditions. With these images the corresponding \emph{PSF} profiles
were also measured together with Eq. (11) and shown in Fig. 8.a).
The \emph{PSF} profiles show a pattern as expected: the \emph{PSF}
of the PPWG shows a sharper and a slightly greater magnitude than
the image acquired without the PPWG. From these profiles, we can infer
that the PPWG did not cause any geometric distortion in the image
and, that the system performance is not affected by the waveguide
despite that the image was obtained far away from the coil array.
Images were again obtained with standard imaging protocols and pulse
sequences in a clinical imager. Uniformity profiles were also computed
using Fig. 8.b) and c) images and shown in Fig. 8.d). Both profiles
have a pretty similar behavior and magnitude. The SNR values of both
experiments were: SNR\textsubscript{PPEWG}/SNR\textsubscript{coil array}=16.74/14.53.
This is a rather interesting result taking into account that the PPWG
images were acquired 100 cm away from the coil array and outside the
MR imager.\bigskip{}

\begin{minipage}[t]{1\columnwidth}%
\noindent \begin{center}
\includegraphics[scale=0.2]{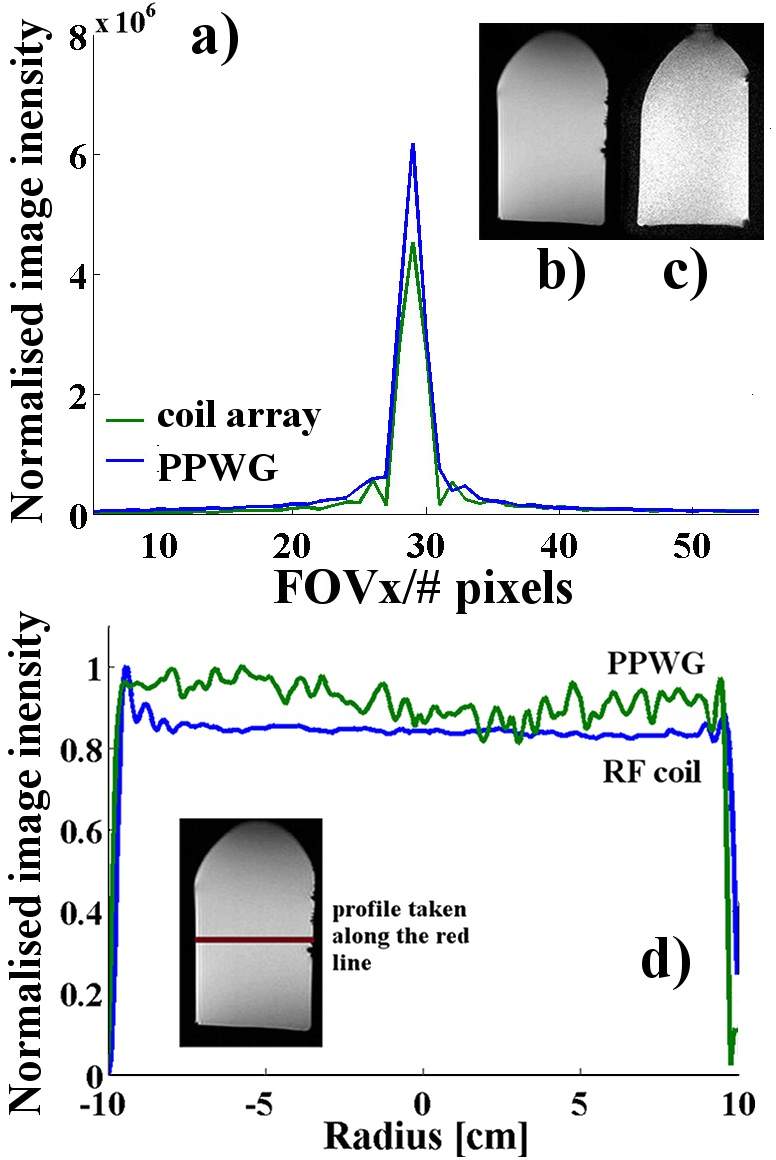}
\par\end{center}

\begin{singlespace}
\noindent \begin{center}
{\footnotesize Figure 8. Profiles of the PSF for the coil array and
PPWG+coil array (a) were computed from phantom images (b) and c))
to measure performance of the 3T imager. A moderately better performance
is shown by the twMRI experiment. With these image data, uniformity
profiles were also computed (d) showing a quite similar performance
despite the fact that the coil array was 1 m away.}
\par\end{center}\end{singlespace}
\end{minipage}\bigskip{}

T1-weighted images of a healthy volunteer\textquoteright{}s entire
leg were obtained using: a) the whole-body coil, b) the PPWG and c)
a pair of RF coils are shown in Fig. 9. As shown, good quality images
were acquired with this simple waveguide and the two circular-shape
flexible coils. The FOV used to cover the entire leg is much larger
than the coil size (11 cm diameter), and the B1 field is not attenuated
along the entire leg; moreover at the waveguide ends, a signal increase
is observed. An extra 5 cm can also be gained with the waveguide scheme.
It is also important to highlight that one single FOV covering the
entire leg was applied such that digital processing was not necessary
as reported in {[}16{]}. Additionally, the leg image acquired with
the pair of single RF coils (Fig. 9.c) shows really poor quality image.
The regions near the two surface coils show very good contrast and
a good image quality compared to the central region. Nevertheless,
an increment in the noise level is also observed in the waveguide
image (Fig. 9.b)) for the central region; this is likely due to the
fold over effect caused by a FOV whose size is greater than the common
FOV sizes used in clinical practice and, no fold over suppression
was applied. \bigskip{}

\begin{minipage}[t]{1\columnwidth}%
\noindent \begin{center}
\includegraphics[scale=0.4]{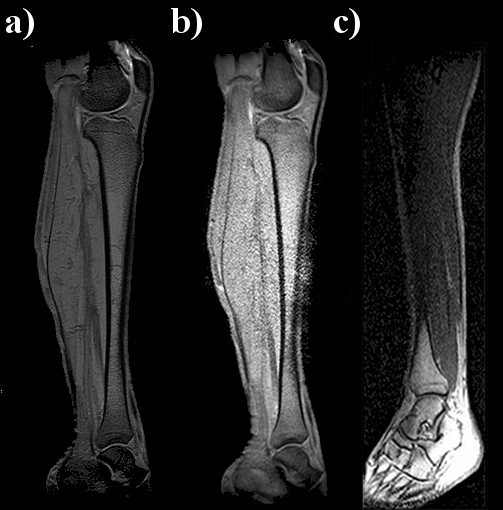}
\par\end{center}

\begin{singlespace}
\noindent \begin{center}
{\footnotesize Figure 9. T1-weighted images of a human leg acquired
with the whole-body birdcage coil (a) and PPWG + pair of RF coils
(b) and the pair of coils only (c). The image acquired with the twMR
approach shows an increment of noise at the centre of the leg not
present in the other two leg images. The image obtained with the pair
of coil shows a really poor signal compared to the other images. An
evident signal contrast between the central and the foot regions can
be appreciated: the higher image intensity can be explained by the
close proximity of the coil to the foot (Fig. 1.c). }
\par\end{center}\end{singlespace}
\end{minipage}\bigskip{}

Another possible source of error is the image reconstruction scheme
which is not the best suited for twMRI experiments. This is a matter
of concern and deserves a thorough investigation. However, the performance
of the travelling-wave approach shows a great potential for imaging
large FOVs. The images in Fig. 9.a) and b) were normalized to the
standard deviation of the image for comparison purposes and shown
in Fig. 10. Comparison plots of image intensity distributions were
also computed using the image data from Fig. 9 for both images. Fig.
9.c) image was shows very poor image quality and it was not used in
the image comparison. 

\bigskip{}

{\small }%
\begin{minipage}[t]{1\columnwidth}%
\noindent \begin{center}
\includegraphics[scale=0.8]{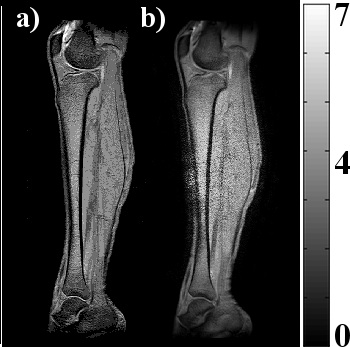}
\par\end{center}

\begin{singlespace}
\noindent \begin{center}
{\footnotesize Figure 10. Comparison of normalized images acquired
with the whole-body birdcage coil (a) and PPWG + pair of RF coils
(b). Central region of (b) image is greater that the other leg image.
Noise is probably due to the fold over effect and that no fold over
suppression was applied. }
\par\end{center}\end{singlespace}
\end{minipage}{\small \par}

\bigskip{}
Fig. 11 shows the distribution forms for both directions. The waveguide-acquired
profile and the birdcage coil-acquired image show a pretty similar
performance. The Fig. 4.b) simulation agrees very well with the experimental
results shown in Fig. 11.b) for the PPWG-acquired image. In our study,
the experimental and numerical results show that the restriction of
a 7T/60 cm system is removed, allowing this approach to be used in
a wider variety of applications involving imagers with much smaller
magnet bores. The PPWG most likely has the simplest waveguide configuration
available, allowing the travelling-wave approach to be implemented
easily. It is much easier to build a waveguide like this and use standard
surface RF coils to excite travelling-waves in a magnet bore. We suspect
that by operating the coil array in the transceiver mode, the image
SNR will be improved. 

\bigskip{}

\noindent \begin{center}
{\small }%
\begin{minipage}[t]{1\columnwidth}%
\begin{center}
\includegraphics[scale=0.3]{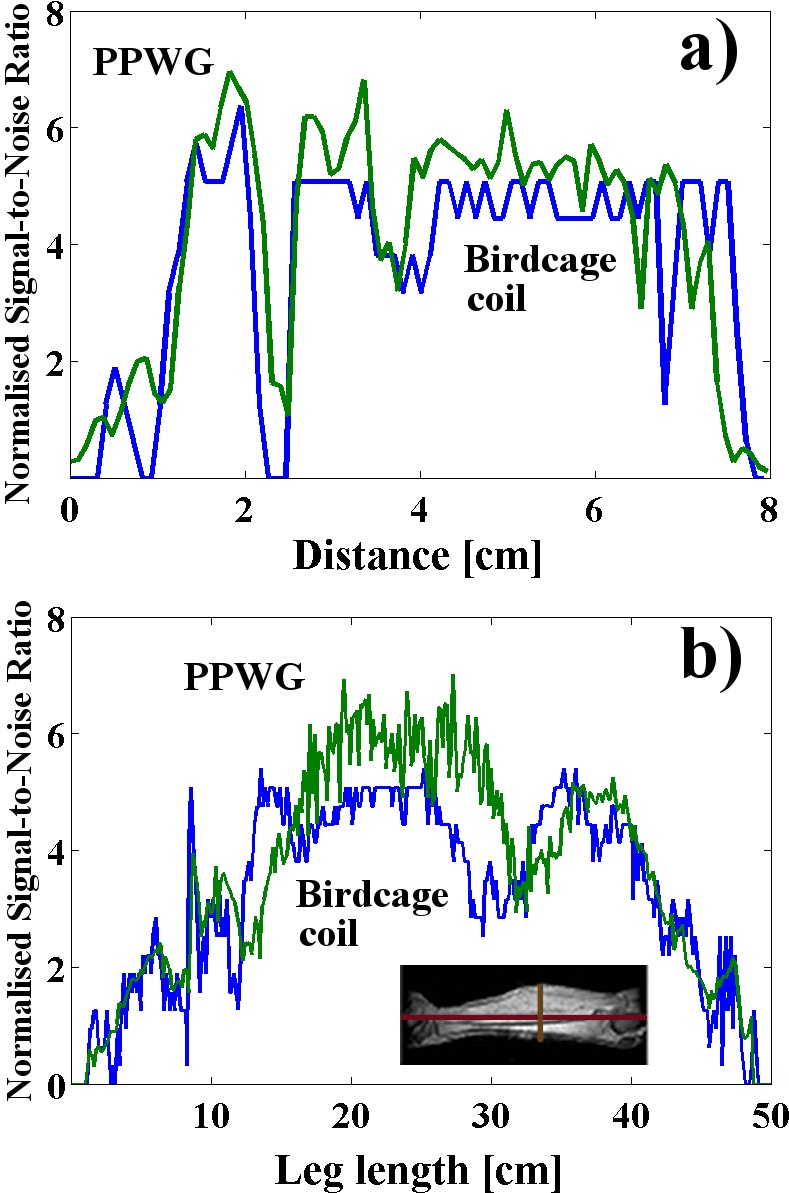}
\par\end{center}

\begin{singlespace}
\noindent \begin{center}
{\footnotesize Figure 11. Comparison plots of normalized SNR distributions
of Fig. 10 images, profiles were taken across the brown line (a) and
along the red line (b) in the leg image. These profiles show a pretty
similar pattern and performance. A strong concordance can be observed
in (b) PPWG profiles with the B1 simulation profiles of Fig. 5.a).}
\par\end{center}\end{singlespace}
\end{minipage}\bigskip{}

\par\end{center}

\section{Conclusion }

We have experimentally demonstrated that the use of PPWGs can produce
good SNR images for relatively large fields of view. We have shown
that the waveguide approach can also be used with magnetic field intensities
lower than 7 T and whole-body systems. Overcoming the limitation of
a cutoff frequency, as demonstrated in this work, provides further
freedom of implementation in a variety of geometries and for multinuclear
operation, or for measuring several or extended samples along a waveguide.
Further investigation is required to explain the physical mechanisms
involved in the RF signal with a dielectric object inside a waveguide
and its implications on image quality. A natural step ahead is to
extend this approach to acquire images of the entire human body. These
results pave the way to a further implementation of the travelling-wave
approach in other applications using lower magnetic field intensities
with bores of different dimensions.

\section*{Acknowledgments }

We would like to acknowledge financial funding from CONACYT-Mexico
under grant no. 166404 and Ph. D. scholarships.


\begin{thebibliography}{References}
\bibitem{key-1}Brunner DO, Zanche ND, Frohlich J, Paska J, Pruessmann
KP. Travelling-wave nuclear magnetic resonance. Nature. 2009;457:994\textendash{}999. 

\bibitem{key-2}Brunner DO, Pruessmann KP. Reciprocity Relations in
Travelling Wave MRI. Proc. Intl. Soc. Mag. Reson. Med. 2009;17:2943. 

\bibitem{key-3}Mueller M, Alt S, Umathum R, Semmler W, Bock M. Targeted
Traveling Wave MRI. Proc. Intl. Soc. Mag. Reson. Med. 2010;18:428. 

\bibitem{key-4}Brunner DO, Paska J, Froehlich J, Pruessmann KP. Traveling-Wave
RF Shimming Parallel MRI. Mag. Reson. Med. 2011;66:290-300. 

\bibitem{key-5}Brenner D, Geschewski F, Felder J, Vahedipour K, Shah
NJ. Experimental Verification of Numerical EM Field Simulations for
Ultra-High Field Travelling Wave MRI. Proc. Intl. Soc. Mag. Reson.
Med. 2011;19:1907.

\bibitem{key-6}Geschewski F, Felder J, Shah N. Optimum Coupling of
Travelling Waves in a 9.4T Whole- Body Scanner. Proc. Intl. Soc. Mag.
Reson. Med. 2010;18:1468. 

\bibitem{key-7}F. Vazquez, R. Martin, O. Marrufo, A. O. Rodriguez,
Waveguide Magnetic Resonance Imaging at 3 Tesla. Proc. Intl. Soc.
Mag. Reson. Med. 2010;18:6484. 

\bibitem{key-8}Cheng DK. Field and Wave Electromagnetics, 2nd Ed.
Addison-Wesley, Reading, 1983. pp. 456-461. 

\bibitem{key-9}De Flaviis F. Guided Waves, in The Electrical Engineering
Handbook, W.-K. Chen, Elsevier Academic Press, Burlington, 2005, pp.
539-551. 

\bibitem{key-10}Jin J. The Finite Element Method in Electromagnetics.
New York: John Wiley \& Sons, Inc., 2002. 

\bibitem{key-11}Deibel JA, Escarra M, Berndsen N, Wang K, Mittleman
DM. Finite-element method simulations of guided wave phenomena at
teraHertz frequencies. Proc. IEEE. 2007;95:1624-1640. 

\bibitem{key-12}RF Module Model Library, COMSOL: Burlington, 2008.
p. 117. 

\bibitem{key-13}Fisher HL, Snyder WS. Distribution of dose in the
body from a source of gamma rays distributed uniformly in an organ.
Report No. ORNL-4168, Oak Ridge National Laboratory, Oak Ridge, TN.
1967. 

\bibitem{key-14}Kerr GD, Hwang JM, Jones RM. A mathematical model
of a phantom developed for use in calculations of radiation dose to
the body and major internal organs of a Japanese adult. Report No.
ORNL/TM-5336, Oak Ridge National Laboratory, Oak Ridge, TN. 1976. 

\bibitem{key-15}F. Vazquez, O. Marrufo, A. O. Rodriguez. Simple method
for B1 mapping at 7 Tesla, Proc. Eur. Soc. Mag. Reson. Med. Biol.
2011;589:143.

\bibitem{key-16}Webb AG, Collins CM, Versluis MJ, Kan HE, Smith NB.
MRI and localized proton spectroscopy in human leg muscle at 7 tesla
using longitudinal traveling waves. Magn. Reson. Med. 2010;63:297\textendash{}302. \end{thebibliography}
\end{document}